# Large Thermoelectric Power Factor at Low Temperatures in One-Dimensional Telluride $Ta_4SiTe_4$

Takumi Inohara,[1] Yoshihiko Okamoto,[1,3,a)] Youichi Yamakawa,[2,3] Ai Yamakage,[1,3] and Koshi Takenaka[1]

[1]*Department of Applied Physics, Nagoya University, Nagoya 464-8603, Japan*
[2]*Department of Physics, Nagoya University, Nagoya 464-8602, Japan*
[3]*Institute for Advanced Research, Nagoya University, Nagoya 464-8601, Japan*

We report the discovery of a very large thermoelectric power over –400 μV K$^{-1}$ in the whisker crystals of a one-dimensional telluride $Ta_4SiTe_4$, while maintaining a low electrical resistivity of ρ = 2 mΩ cm, yielding a very large power factor of P = 80 μW cm$^{-1}$K$^{-2}$ at an optimum temperature of 130 K. This temperature is widely controlled from the cryogenic temperature of 50 K to room temperature by chemical doping, resulting in the largest P of 170 μW cm$^{-1}$K$^{-2}$ at 220–280 K. These P values far exceed those of the $Bi_2Te_3$-$Sb_2Te_3$ alloys at around room temperature, offering an avenue for realizing the practical-level thermoelectric cooling at low temperatures. The coexistence of a one-dimensional electronic structure and a very small band gap appearing in the vicinity of the Dirac semimetals probably causes the very large power factors in $Ta_4SiTe_4$, indicating that the "one-dimensional Dirac semimetal" is a promising way to find high-performance thermoelectric materials for the low temperature applications.

Thermoelectric cooling, which involves solid-state refrigeration by conduction electrons, is a promising candidate for the next-generation of refrigeration technologies in replacing vapor compression cooling using gaseous refrigerants such as CFCs. The thermoelectric cooler can be downsized without a decrease of the conversion efficiency between thermal and electric energies, which is a strong advantage for local cooling of electronic devices. At present, Peltier devices using $Bi_2Te_3$-based materials have been put to practical use in the cooling of infrared sensors and the temperature control of laser diodes at around room temperature. Practical cooling at much lower temperatures enables us to refrigerate cryogenic electronic devices, such as superconducting sensors using cuprates, resulting in much wider applications of thermoelectric cooling.[1-3]

However, there is currently no bulk material with a high enough thermoelectric efficiency to reach a practical level in the low temperature region below 250 K. To achieve such high efficiency, improving the power factor, $P = S^2/\rho$, of a material, where S and ρ are the thermoelectric power and electrical resistivity, respectively, is the first step. A target value of P is ~35 μW cm$^{-1}$K$^{-2}$, which is the typical value of $Bi_2Te_3$-based practical materials.[4] In this case, as shown in Table I, $Bi_2Te_3$-based materials show large S of typically 240 μV K$^{-1}$, which is essential for achieving high thermoelectric efficiency.[1,5] A number of candidate materials for low temperature applications, such as $Bi_{1-x}Sb_x$ (x = 0.15),[6] $CsBi_4Te_6$,[7,8] $ZrTe_{4.95}Sb_{0.05}$,[9] $CePd_3$,[10] and $YbAl_3$,[11] have been developed thus far. As seen in Table I, however, their S values are smaller than those of $Bi_2Te_3$-based materials, except for that of $FeSb_2$, which shows an exceptionally large S at 10 K due to the phonon drug effect,[12,13] resulting in their thermoelectric efficiency remaining below the practical level. Discovering promising materials that exhibit the large thermoelectric power of |S| > 200 μV K$^{-1}$ at low temperatures in keeping small electrical resistivity is an important step for thermoelectric cooling to be used in a wide range of applications.

In this study, we focus on the thermoelectric properties of $Ta_4SiTe_4$. $Ta_4SiTe_4$ was first synthesized by Badding and DiSalvo in a needle-like crystalline form.[14] This compound crystallizes in the orthorhombic structure with the space group *Pbam* (No. 55), as shown in Fig. 1(a).[14] Each Si atom is coordinated to a $Ta_8$ square antiprism, which is further coordinated to eight Te atoms in the same way. The $SiTa_8Te_8$ nested antiprisms form a $Ta_4SiTe_4$ chain along the c axis, which is the needle direction, by sharing their basal planes. The chains are weakly bonded by van der Waals forces and form a nearly close-packed structure, where the distances between neighboring chains differ by only 0.1%. $Ta_4SiTe_4$ was predicted to be metallic by first-principles calculations and the molecular orbital theory.[15] In contrast, the electrical resistivities of the whisker crystals show semiconducting behavior with a very small energy gap,[16,17] although metallic behavior is also reported in the whisker

---

a) Electronic mail: yokamoto@nuap.nagoya-u.ac.jp

TABLE I. Maximum power factor $P_{max}$ and thermoelectric parameters of the typical thermoelectric cooling materials.

| | $P_{max}$ ($\mu$W cm$^{-1}$K$^{-2}$) | $T_{max}$ (K) | $S(T = T_{max})$ ($\mu$V K$^{-1}$) | $\rho(T = T_{max})$ (m$\Omega$ cm) | $ZT_{max}$ | Reference |
|---|---|---|---|---|---|---|
| Ta$_4$SiTe$_4$ | 80 | 130 | −410 | 2.1 | | This work |
| (Ta$_{0.999}$Mo$_{0.001}$)$_4$SiTe$_4$ | 170 | 240 | −310 | 0.58 | | This work |
| Bi$_2$Te$_3$ | 35 | 300 | 240 | 1.7 | 1.0 | 4 |
| Bi$_{0.85}$Sb$_{0.15}$ | 150 | 77 | −150 | 0.15 | 0.5 | 6 |
| CsBi$_4$Te$_6$ [a] | 52 | 184 | 163 | 0.52 | 0.8 | 8 |
| ZrTe$_{4.95}$Sb$_{0.05}$ | 30 | 150 | 190 | 1.2 | | 9 |
| CePd$_3$ | 92 | 140 | 125 | 0.17 | 0.03 | 10 |
| YbAl$_3$ | 220 | 120 | −75 | 0.025 | | 11 |

[a] 0.05% SbI$_3$ doped sample.

crystals prepared by another group.[18] In this study, we report thermoelectric properties of Ta$_4$SiTe$_4$ and its substituted compounds. They show very large thermoelectric power far exceeding $|S| = 200$ $\mu$V K$^{-1}$ in the wide temperature region between 50 K and room temperature, in keeping the low resistivity enough for the practical-level cooling. This result indicates that Ta$_4$SiTe$_4$ is a highly promising material for the thermoelectric cooling in the low temperature region.

The whisker crystals of Ta$_4$SiTe$_4$ and its substituted compounds were synthesized by the crystal growth in a vapor phase. A mixture of a $2 - 2x : 2x : 1 : 2y : 2 - 2y$ molar ratio of Ta (Rare Metallic, 99.9%), Mo (Kojundo Chemical Laboratory, ≥99.9%), Si (Kojundo Chemical Laboratory, ≥99.9%), and Sb powders (Kojundo Chemical Laboratory, 99.9%) and Te chips (Rare Metallic, 99.999%), which is 1 g in total, was sealed in an evacuated quartz tube with 10 mg of TeCl$_4$ powder. The tube was heated to and kept at 873 K for 24 h, 1423 K for 96 h, and then furnace cooled to room temperature. The whisker crystals, with typically several mm long and several $\mu$m in diameter as shown in Fig. 1(b), were picked from the obtained samples. The Mo content $x$ and Sb content $y$ of the whisker crystals described below are the nominal values. Sintered samples of Ta$_4$SiTe$_4$ were prepared by a solid-state reaction method. A $4 : 1–1.1 : 4$ molar ratio of Ta powder (Rare Metallic, 99.9%), Si powder (Kojundo Chemical Laboratory, ≥99.9%), and Te chips (Rare Metallic, 99.999%) were mixed and sealed in a quartz tube with 0.04–0.05 MPa of Ar gas. The tube was heated to and kept at 1323 K for 48 h and then furnace cooled to room temperature. The obtained sample was pulverized, pressed into pellets, and then sealed in a quartz tube with 0.04–0.05 MPa of Ar gas. The tube was heated at 1173 K for 3 h and then furnace cooled to room temperature.

Sample characterization was performed by powder X-ray diffraction analysis for pulverized whisker crystals and sintered samples with Cu K$\alpha$ radiation at room temperature using a RINT-2100 diffractometer. We confirmed that the whisker crystals and the sintered samples are single phase. The electrical resistivity, thermoelectric power, and Hall resistivity measurements were performed using a Physical Property Measurement System (PPMS, Quantum Design) and a home-built apparatus. Thermal conductivity was measured by a standard four-contact method. First principles calculations for Ta$_4$SiTe$_4$ were performed using the WIEN2k code.[19] Experimental structural parameters were used for the calculations.[14]

Figures 1(c) and 1(d) show the temperature dependences of $\rho$ and $S$ of whisker crystals of (Ta$_{1-x}$Mo$_x$)$_4$Si(Te$_{1-y}$Sb$_y$)$_4$ ($x$ = 0–0.05 and $y$ = 0, 0.05) measured along the $c$ axis, respectively. The $\rho$ of Ta$_4$SiTe$_4$ increases with decreasing temperature from room temperature and shows a broad peak at around 200 K. With further decreasing temperature, $\rho$ exhibits a local minimum of approximately 2 m$\Omega$ cm at around 100 K and then strongly increases below 50 K. This temperature dependence is consistent with the results reported in Refs. 16 and 17. The Sb-doped sample ($y$ = 0.05), which is the hole-doped sample of Ta$_4$SiTe$_4$, shows larger $\rho$ than that of the undoped one, as shown in Fig. 1(c). In contrast, only 0.1% Mo doping to the Ta site, which is electron doping of Ta$_4$SiTe$_4$, drastically decreases $\rho$ and changes the temperature dependence into metallic behavior with $d\rho/dT > 0$ toward the lowest measured temperature of 2 K. The $\rho$ further decreases with increasing the Mo content $x$, resulting in $\rho$ of ~0.1 m$\Omega$ cm at room temperature in the $x$ = 0.05 sample. Except for the strong increase below 50 K for undoped Ta$_4$SiTe$_4$ and $y$ = 0.05 samples, the $\rho$ values of all measured samples are less than 5 m$\Omega$ cm, which is small enough for a practical thermoelectric material.

The thermoelectric powers of (Ta$_{1-x}$Mo$_x$)$_4$Si(Te$_{1-y}$Sb$_y$)$_4$ samples are negative, as shown in Fig. 1(d), indicating that the electron carriers are dominant in this system. Undoped Ta$_4$SiTe$_4$ shows $S \sim -200$ $\mu$V K$^{-1}$ at room temperature. The $|S|$ of Ta$_4$SiTe$_4$ strongly increases with decreasing temperature, reaching a quite large value of $|S| = 400$ $\mu$V K$^{-1}$ at 100–200 K, followed by a strong decrease toward zero below 50 K. The $S$ of the $y$ = 0.05 sample, a hole-doped sample, exhibits a similar temperature dependence to that of the undoped sample and becomes a larger value of $|S| = 500$ $\mu$V K$^{-1}$ between 30 and 180 K. In contrast, electron doping by Mo substitution results in a systematic decrease of the maximum value of $|S|$, as seen in Fig. 1(d). The $|S|$ of the $x$ = 0.001 sample reaches 300 $\mu$V K$^{-1}$ at around 250 K, which is extremely large for the metallic compound. The thermoelectric powers of the $x \geq 0.01$ samples show metallic temperature dependences in the measured temperature region, where $|S|$ increases with increasing temperature. The $x$ = 0.05 sample, which is the most heavily electron-doped sample in this study, shows $|S|$ = 75 $\mu$V K$^{-1}$ at room temperature, which is still large for the metallic system.

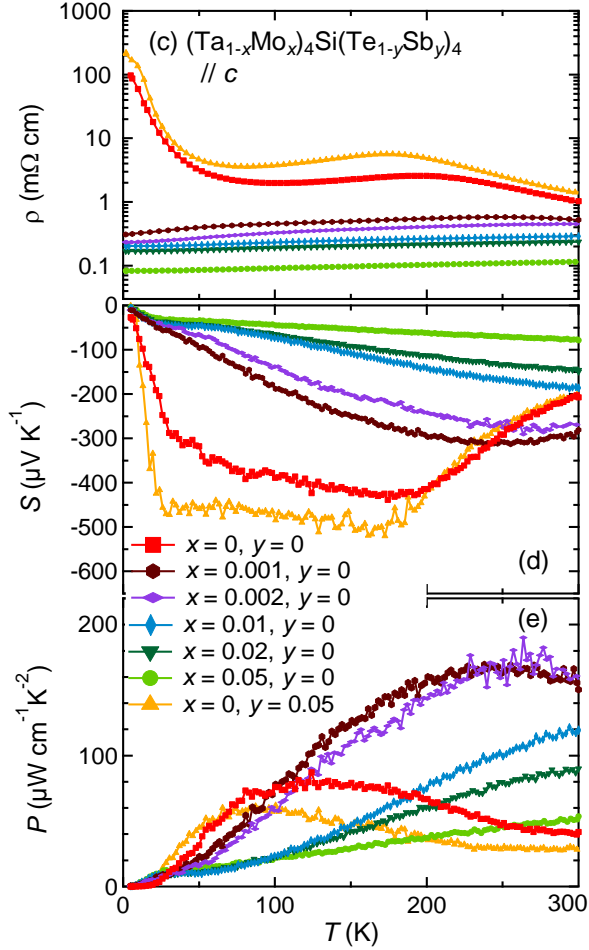

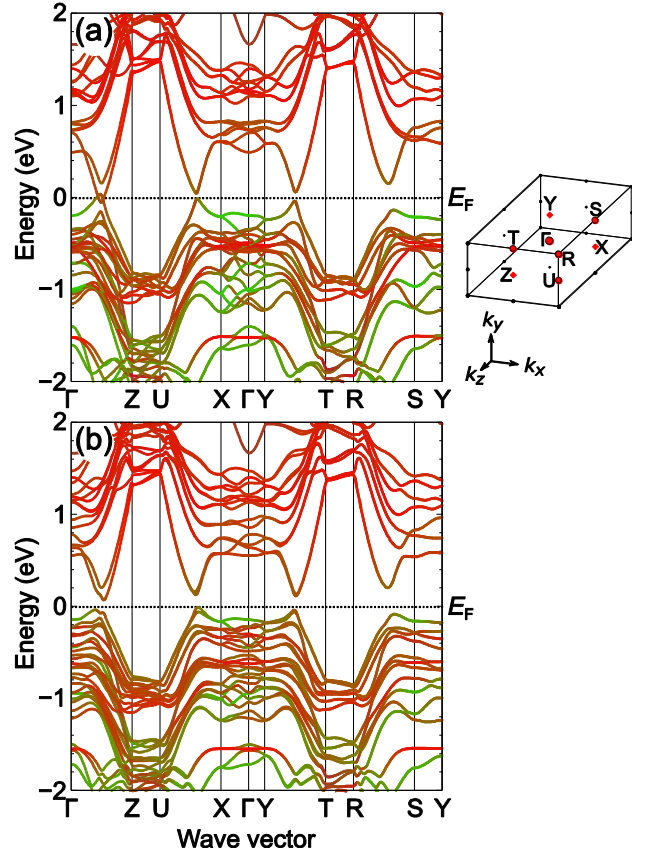

FIG. 1. (a) Crystal structure of $Ta_4SiTe_4$. The orthorhombic unit cell is indicated by broken lines. A $Ta_4SiTe_4$ chain is shown in the right side. (b) A scanning electron microscope image of a $Ta_4SiTe_4$ whisker crystal. (c) Electrical resistivities, (d) thermoelectric powers, and (e) power factors of the $(Ta_{1-x}Mo_x)_4Si(Te_{1-y}Sb_y)_4$ ($x$ = 0–0.05 and $y$ = 0, 0.05) whisker crystals measured along $c$ axis.

Thus, the $(Ta_{1-x}Mo_x)_4Si(Te_{1-y}Sb_y)_4$ whisker crystals almost always show small $\rho$ comparable to those of the practical materials, while can show very large $|S|$ substantially larger than 200 $\mu$V K$^{-1}$, resulting in very large $P$ shown in Fig. 1(e). The undoped $Ta_4SiTe_4$ shows a maximum value of the power factor of $P_{max}$ = 80 $\mu$W cm$^{-1}$K$^{-2}$ at the optimum temperature of $T_{max}$ = 130 K. With increasing $x$, $T_{max}$ is shifted to the higher temperature. The $x$ = 0.001 and 0.002 samples show a surprisingly large $P_{max}$ of 170 $\mu$W cm$^{-1}$K$^{-2}$ at $T_{max}$ = 240 and 260 K, respectively. As shown in Table I, these $P_{max}$ values are almost twice and four times larger than $P_{max}$ = 35 $\mu$W cm$^{-1}$K$^{-2}$ of the $Bi_2Te_3$-based materials,[4,20] respectively, indicating that the $P_{max}$ of this system far exceeds the level of practical materials in a wide temperature region between 50 K and room temperature.

FIG. 2. Electronic structures without (a) and with (b) spin-orbit coupling. The colors red and green represent the contributions of the $d$-orbital of Ta and $p$-orbital of Te, respectively. The Fermi level is set to 0 eV. The first Brillouin zone is shown in the right panel of (a).

We now discuss the origin of the very large $P$ realized in the low temperature regions of $Ta_4SiTe_4$ and its substituted compounds. As seen in Fig. 1(d), $Ta_4SiTe_4$ shows a very large $|S|$ of approximately 400 $\mu$V K$^{-1}$ at 100–200 K. In this temperature region, $|S|$ and $\rho$ gradually increase with increasing temperature, resulting in the $P_{max}$ of ~80 $\mu$W cm$^{-1}$K$^{-2}$. Although Mo or Sb doping varies the $T_{max}$, it is common for all samples that the large $|S|$ appears in the temperature region with $d\rho/dT > 0$.

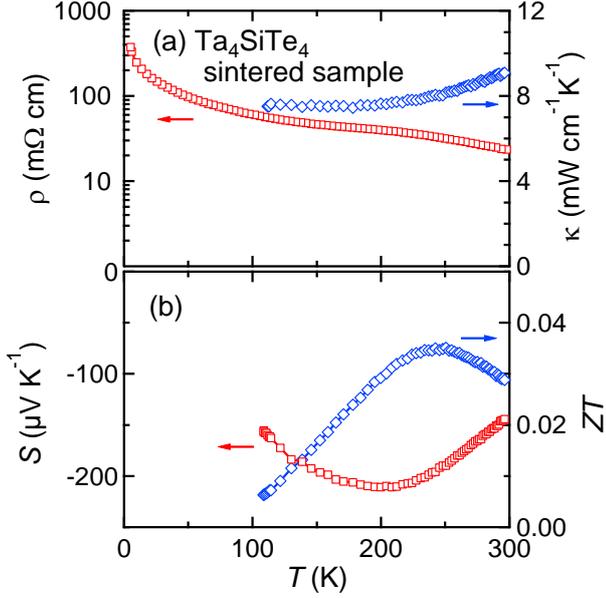

FIG. 3. Temperature dependences of electrical resistivity, thermal conductivity (a), thermoelectric power, and dimensionless figure of merit (b) of a $Ta_4SiTe_4$ sintered sample.

The large $|S|$ in the low temperature region of $Ta_4SiTe_4$ is supposed to be closely related to the one-dimensional electronic structure. As shown in Fig. 1(a), $Ta_4SiTe_4$ crystallizes in the strongly one-dimensional structure, where $Ta_4SiTe_4$ chains lying along the $c$ axis are loosely bonded by van der Waals forces. In fact, as seen in the first principles calculation results shown in Fig. 2(a) and 2(b), band dispersions along the $k_x$ and $k_y$ directions are almost flat and much weaker than that along the $k_z$ direction. This results in that the effective mass, $m^*$, in the $k_x$ and $k_y$ directions being much larger than that in the $k_z$ direction. Such an anisotropic $m^*$ enhances the $|S|$, in keeping with the small $\rho$ along the $c$ axis, giving rise to a large $P$, as noted in the studies of $CsBi_4Te_6$.[21,22] This type of an anisotropic metallic state is probably realized and enhances $S$ in the $d\rho/dT > 0$ temperature regions of $Ta_4SiTe_4$ and its substituted compounds, because $d\rho/dT > 0$ indicates the presence of a coherent electronic transport not only along the $Ta_4SiTe_4$ chains but also in the interchain direction, even though it seems to be weak. In the higher temperature region with $d\rho/dT < 0$, the electronic transport in the interchain direction might become incoherent, meaning that the concept of the anisotropic metallic state is not applicable, resulting in the smaller $|S|$.

Another important feature of $Ta_4SiTe_4$ is that this type of one-dimensional electronic state realizes in the system with a very small band gap. As shown in Fig. 2(b), which is the band structure of $Ta_4SiTe_4$ calculated with the spin-orbit interaction, a band gap of $\Delta \sim 0.1$ eV opens at the Fermi energy $E_F$. It is known that there is a positive correlation between $\Delta$ and $T_{max}$, for example, $\Delta = 0.25$ eV and $T_{max} = 300$ K for $Bi_2Te_3$, while $\Delta = 0.5$ eV and $T_{max} = 500$ K for $PbTe$.[4,23] $\Delta \sim 0.1$ eV of $Ta_4SiTe_4$ seems to be consistent with $T_{max} = 130$ K. In fact, the electron carrier density, $n$, of the $Ta_4SiTe_4$ sintered sample is estimated to be $1 \times 10^{19}$ cm$^{-3}$ at 200 K by Hall resistivity experiments. This $n$ is similar to those of the practical materials such as $Bi_2Te_3$ and $PbTe$.[4] $n$ increases with increasing temperature and becomes $3 \times 10^{19}$ cm$^{-3}$ at 300 K, which is considerably larger than the $n$ for the practical materials. These results indicate that the $n$ of $Ta_4SiTe_4$ is appropriately tuned as a thermoelectric cooling material in the low temperature region.

We believe that the very small band gap realized in $Ta_4SiTe_4$ is closely related to an aspect of this compound as a Dirac fermion system. As seen in the band structure calculated without the spin-orbit interaction shown in Fig. 2(a), the valence and conduction bands cross on the $\Gamma$–$Z$ line at around $E_F$. This band crossing point seems to form rings in the $k_y$–$k_z$ and $k_z$–$k_x$ planes surrounding a point on the $\Gamma$–$Z$ line, suggesting that $Ta_4SiTe_4$ is classified as a line-node Dirac semimetal.[24,25] The very small gap of $\Delta \sim 0.1$ eV displayed in Fig. 2(b) is a spin-orbit gap caused by the strong spin–orbit interaction of Ta 5$d$ and Te 5$p$ electrons. Thus, $Ta_4SiTe_4$ and its substituted compounds are supposed to show the unprecedentedly large $P$ at low temperatures due to the fact that the strongly one-dimensional electronic state is realized in a very narrow gap system appearing in the vicinity of the Dirac fermion system. Such a "one-dimensional Dirac semimetal" can be a promising avenue to develop high-performance thermoelectric materials for the low temperature applications.

Finally, we note the dimensionless figure of merit $ZT$. Estimating $ZT = PT/\kappa$ by combining $P$ and thermal conductivity $\kappa$ is important for the evaluation of the thermoelectric energy conversion efficiency. However, we cannot measure the $\kappa$ of the whisker crystals prepared in this study, because they are too thin. As shown in Fig. 3, a sintered sample of $Ta_4SiTe_4$ shows a maximum $ZT$ value of 0.035, much smaller than those of the $Bi_2Te_3$-based practical materials (Table I). This is mainly because the $\rho$ value of the sintered sample is 30–40 times larger than that of the whisker crystal measured along the $c$-axis, reflecting the strongly one-dimensional electronic structure in this system. Alternatively, the $\kappa$ value of a $Ta_4SiTe_4$ sintered sample is 9 mW cm$^{-1}$ K$^{-1}$ at room temperature, as shown in Fig. 3(a), almost all of which are the phonon contribution $\kappa_{lat}$. This value is quite small comparable to those of amorphous or glassy materials. If we use the thermal conductivity of the conduction electrons, $\kappa_{el}$, of the whisker crystals, obtained by applying Wiedemann–Franz law to the $\rho$ data shown in Fig. 1(c), and the $\kappa_{lat}$ of the sintered samples, the upper limits of $ZT$ of the whisker crystals are estimated to be 2.2 for $(Ta_{0.999}Mo_{0.001})_4SiTe_4$ at 250 K and 1.4 for $Ta_4SiTe_4$ at 180 K. Considering the facts that $Ta_4SiTe_4$ has a complex crystal structure, mainly comprises heavy elements, and contains Te atoms with a wide distribution of mass numbers of the natural isotopes,[4,26] the small $\kappa_{lat}$ of the sintered sample is likely to reflect not only the extrinsic factors such as grain boundaries but also intrinsic ones. This discussion strongly suggests that the whisker crystals can show low $\kappa$, possibly resulting in large $ZT$ exceeding those of the practical materials, although the bundles of the whisker crystals or highly-oriented polycrystalline samples need to be prepared for the practical use of this material.

In conclusion, one-dimensional telluride $Ta_4SiTe_4$ and its substituted compounds are found to show high thermoelectric performance indicated by the very large thermoelectric power factors of up to 170 $\mu W\ cm^{-1} K^{-2}$. They appear in the low temperature region and are accompanied by very large $S$ over $-200\ \mu V\ K^{-1}$, indicating that $Ta_4SiTe_4$ is a strong candidate material for the low temperature applications of thermoelectric cooling. The high thermoelectric performance of $Ta_4SiTe_4$ is mostly caused by the coexistence of strongly one-dimensional electronic structure and a very small band gap appearing in the vicinity of the Dirac semimetals.

**ACKNOWLEDGMENTS**

This work was partly supported by JSPS KAKENHI (Grant No. 16K13664 and 16H03848), the Research Foundation for the Electrotechnology of Chubu, the Iketani Science and Technology Foundation, and the Asahi Glass Foundation.

**REFERENCES**

[1] T. M. Tritt and M. A. Subramanian, MRS Bull. **31**, 188 (2006).
[2] R. P. Huebener and C. C. Tsuei, Cryogenics **38**, 325 (1998).
[3] T. Metzger and R. P. Huebener, Cryogenics **39**, 235 (1999).
[4] G. D. Mahan, *Solid State Physics* (Academic Press, New York, USA, 1997) Vol. 51, pp. 81-157.
[5] G. Mahan, B. Sales, and J. Sharp, Phys. Today **50**, 42 (1997).
[6] W. M. Yim and A. Amith, Solid State Electron. **15**, 1141 (1972).
[7] D.-Y. Chung, T. Hogan, P. Brazis, M. Rocci-Lane, C. Kannewurf, M. Bastea, C. Uher, and M. G. Kanatzidis, Science **287**, 1024 (2000).
[8] D.-Y. Chung, T. P. Hogan, M. Rocci-Lane, P. Brazis, J. R. Ireland, C. R. Kannewurf, M. Bastea, C. Uher, and M. G. Kanatzidis, J. Am. Chem. Soc. **126**, 6414 (2004).
[9] R. T. Littleton IV, T. M. Tritt, J. W. Kolis, D. R. Ketchum, N. D. Lowhorn, and M. B. Korzenski, Phys. Rev. B **64**, 121104 (2001).
[10] R. J. Gambino, W. D. Grobman, and A. M. Toxen, Appl. Phys. Lett. **22**, 506 (1973).
[11] H. J. Van Daal, P.B. Van Aken, and K. H. J. Buschow, Phys. Lett. A **49**, 246 (1974).
[12] A. Bentien, S. Johnsen, G. K. H. Madsen, B. B. Iversen, and F. Steglich, EPL **80**, 17008 (2007).
[13] H. Takahashi, R. Okazaki, S. Ishiwata, H. Taniguchi, A. Okutani, M. Hagiwara, and I. Terasaki, Nat. Commun. **7**, 12732 (2016).
[14] M. E. Badding and F. J. DiSalvo, Inorg. Chem. **29**, 3952 (1990).
[15] J. Li, R. Hoffmann, M. E. Badding, and F. J. DiSalvo, Inorg. Chem. **29**, 3943 (1990).
[16] M. E. Badding, R. L. Gitzendanner, R. P. Ziebarth, and F. J. DiSalvo, Mat. Res. Bull. **29**, 327 (1994).
[17] K. Ahn, T. Hughbanks, K. D. D. Rathnayaka, and D. G. Naugle, Chem. Mater. **6**, 418 (1994).
[18] A. Stolovits, A. Sherman, K. Ahn, and R. K. Kremer, Phys. Rev. B **62**, 10565 (2000).
[19] P. Blaha P., K. Schwarz, G. Madsen, D. Kvasnicka, and J. Luitz, *WIEN2k, an augmented plane wave + local orbitals program for calculating crystal properties* (Tech. Universität Wien, Vienna, 2001).
[20] L. R. Testardi, J. N. Bierly Jr., and F. J. Donahoe, J. Phys. Chem. Solid **23**, 1209 (1962).
[21] V. A. Greanya, W. C. Tonjes, R. Liu, C. G. Olson, D.-Y. Chung, and M. G. Kanatzidis, Phys. Rev. B **65**, 205123 (2002).
[22] P. Larson, S. D. Mahanti, D.-Y. Chung, and M. G. Kanatzidis, Phys. Rev. B **65**, 045205 (2002).
[23] G. D. Mahan, J. Appl. Phys. **65**, 1578 (1989).
[24] L. S. Xie, L. M. Schoop, E. M. Seibel, Q. D. Gibson, W. Xie, and R. J. Cava, APL Mater. **3**, 041504 (2015).
[25] A. Yamakage, Y. Yamakawa, Y. Tanaka, and Y. Okamoto, J. Phys. Soc. Jpn. **85**, 013708 (2016).
[26] G. A. Slack, *Solid State Physics* (Academic Press, New York, USA, 1979), Vol. 34, pp. 1-71.